# Velocity distribution of free particles in the modified special relativity theory

Jian-Miin Liu
Department of Physics, Nanjing University
Nanjing, The People's Republic of China
On leave. E-mail address: liu@mail.davis.uri.edu

The current Lorentz-invariant field theory, whether classical or quantum, has been plagued with the divergence difficulties. Although the renormalization technique was developed and enjoyed great experimental successes, the renormalized quantum field theory can not be considered as final since it is unable to do anything about the important phenomena in particle physics, the mass difference in some particle groups such as proton and neutron, the $\pi$-mesons, the K-mesons, the $\Sigma$-hyperons and the $\Xi$-hyperons. Moreover, it is hard, from the theoretical point of view, to accept renormalizability as a basic principle of physics. The origins of the divergence difficulties lay deep within the conceptual foundations of the theory. Two foundation stones of the current quantum field theory are: the special theory of relativity and the quantum mechanics theory. Actually, there seems to be a bone to pick between these two stone theories. The Lorentz invariance in the special relativity theory recognizes the light speed c constituting a limitation for transport of matter or energy and transmission of information or causal connection, wheras the quantum mechanics theory contains essentially instantaneous quantum connection. In order to resolve the divergence difficulties, and that bone problem as well, we modified the special theory of relativity by keeping its two fundamental postulates, (i) the principle of relativity and (ii) the constancy of the one-way velocity of light in all inertial frames of reference, and alternatively assuming two generalized Finslerian local structures of gravity-free space and time in the usual inertial coordinate system [1].

*   *   *

In 1905, Einstein published his special theory of relativity [2]. He derived the Lorentz transformation between any two usual inertial coordinate systems, which is the kinematical background for the physical principle of Lorentz invariance. The two fundamental postulates (i) and (ii) were stated by Einstein as the basis for his theory. But the special relativity theory in fact employs, besides these two fundamental postulates, another assumption. This other assumption concerns the Euclidean structure of gravity-free space and the homogeneity of gravity-free time in the usual inertial coordinate system $\{x^r,t\}$, r=1,2,3, $x^1$=x, $x^2$=y, $x^3$=z,

$$dX^2=\delta_{rs}dx^r dx^s, \quad r,s=1,2,3, \tag{1a}$$
$$dT^2=dt^2, \tag{1b}$$

everywhere and every time.

Postulates (i) and (ii) and the assumption Eqs.(1) together yield the Lorentz transformation between any two usual inertial coordinate systems [2-5]. Indeed, though the assumption Eqs.(1) was not explicitly articulated, having been considered self-evident, Einstein said in 1907 [2]: "Since the propagation velocity of light in empty space is c with respect to both reference systems, the two equations, $x_1^2+y_1^2+z_1^2-c^2t_1^2=0$ and $x_2^2+y_2^2+z_2^2-c^2t_2^2=0$ must be equivalent." Leaving aside a discussion of whether postulate (i) implies the linearity of transformation between any two usual inertial coordinate systems and the reciprocity of relative velocities between any two usual inertial coordinate systems, we know that the two equivalent equations, the linearity of transformation and the reciprocity of relative velocities exactly lead to the Lorentz transformation.

Some physicists explicitly articulated the assumption Eqs.(1) in their works on the topic. Pauli wrote: "This also implies the validity of Euclidean geometry and the homogeneous nature of space and time." [4]. Fock said: "The logical foundation of these methods is, in principle, the hypothesis that Euclidean geometry is applicable to real physical space together with further assumptions, viz. that rigid bodies exist and that light travels in straight lines." [5].

Conceptually, the principle of relativity means that there exists an equivalance class of the inertial frames of reference, any one of which moves with a non-zero constant velocity relative to any other. Einstein wrote in his Autobiographical Notes [3]: "in a given inertial frame of reference the coordinates mean the results of certain measurements with rigid (motionless) rods, a clock at rest relative to the inertial frame of reference defines a local time, and the local time at all points of space, indicated by synchronized clocks and taken together, give the time of this inertial frame of reference." As defined by Einstein, each of the inertial frames of reference is supplied with motionless, rigid unit rods of equal length and motionless, synchronized clocks of equal running rate. Then in each inertial frame of reference, an observer can employ his own motionless-rigid rods and motionless-synchronized clocks in the so-called "motionless-rigid rod and motionless-synchronized clock" measurement method to measure space and time intervals. By using this "motionless-rigid rod and motionless-synchronized clock" measurement method, the observer in each inertial frame of reference can set up his own usual inertial coordinate system. The postulate (ii) means that the measured one-way velocity of light is the same constant c in every such usual inertial coordinate system.

The "motionless-rigid rod and motionless-synchronized clock" measurement method is not the only one that each inertial frame of reference has. Following Einstein, we imagine for each inertial frame of reference other measurement methods that are different from the "motionless-rigid rod and motionless-synchronized clock" measurement method. By taking these other measurement methods, according to Einstein, an observer in each inertial frame of reference can set up other inertial coordinate systems, just as well as he can set up his usual inertial coordinate system by taking the "motionless-rigid rod and motionless-synchronized clock" measurement method. We call these other inertial coordinate systems the unusual inertial coordinate systems.

The conventional belief in flatness of gravity-free space and time is natural. But question is, in which inertial coordinate system gravity-free space and time directly display their flatness. The special relativity theory recognizes the usual inertial coordinate system, as shown in Eqs.(1). Making a different choice, we pick a unusual inertial coordinate system, say the primed inertial coordinate system, $\{x'^r, t'\}$, r=1,2,3. We assume that gravity-free space and time possess the flat metric structures in the primed inertial coordinate system, and hence, the following local structures in the usual inertial coordinate system,

$$dX^2 = \delta_{rs} dx'^r dx'^s = g_{rs}(y) dx^r dx^s, \quad r,s=1,2,3, \tag{2a}$$
$$dT^2 = dt'^2 = g(y) dt^2, \tag{2b}$$
$$g_{rs}(y) = K^2(y) \delta_{rs}, \tag{3}$$
$$g(y) = (1 - y^2/c^2), \tag{4}$$
$$K(y) = \frac{c}{2y} \sqrt{1 - y^2/c^2} \ln \frac{c+y}{c-y}, \tag{5}$$

where $y = (y^s y^s)^{1/2}$, $y^s = dx^s/dt$, s=1,2,3. The structures specified by metric tensors $g_{rs}(y)$ and $g(y)$ in the assumption Eqs.(2) can be studied and described through the generalized Finsler geometry [6].

Combining this alternative assumption, instead of the assumption Eqs.(1), with the two fundamental postulates (i) and (ii), we formed the modified theory of special relativity. The modified theory involves two versions of the light speed, infinite speed η in the primed inertial coordinate system and finite speed c in the usual inertial coordinate system [1]. Both speeds η and c are invariant in all inertial frames of reference [1]. The modified theory involves the Galilean addition law among primed velocities and the Einstein addition law among usual (Newtonian) velocities (see below). The modified theory also involves the η-type Galilean transformation between any two primed inertial coordinate systems, in which speed η is invariant, and the localized Lorentz transformation between two corresponding usual inertial coordinate systems, where the space and time differentials take places of the space and time variables respectively in the Lorentz transformation [1]. After confirming that all our experimental data are collected and expressed in the usual inertial coordinate system, we have a new physical principle: the η-type Galilean invariance in the primed inertial coordinate system plus the transformation from the primed to the usual inertial coordinate systems [1]. For quantum field theory, all equations of fields are written in the η-type Galilean-invariant manner in the primed inertial coordinate system, all η-type Galilean-invariant calculations including quantizing these fields by use of the canonical quantization method are done in the primed inertial coordinate system, and the obtained results are all transformed to the usual inertial coordinate system and compared to experimental facts there. We showed that the modified theory together with the quantum mechanics theory features a covergent and invariant quantum field theory in full agreement with ecperimental facts [1].

Since of the mentioned characteristics of the modified theory, all experiments [7-9] supporting the existence of the constancy of the usual light speed c, the Einstein addition law of relative usual velocities, and the Lorentz invariance or the local Lorentz invariance also support the modified theory. These experiments can not be used to judge between the special theory of relativity and the modified theory.

Besides, since the validity of relativistic mechanics in the usual inertial coordinate system remains in the modified theory [1], any mechanics experiment can not be used to judge, either. However, in this paper, we propose a new velocity distribution of free particles to test the non-flat structures Eqs.(2) of gravity-free space and time in the usual inertial coordinate system below.

$$* \quad * \quad *$$

As we define a new type of velocity (primed velocity), $y'^s = dx'^s/dt'$, s=1,2,3, in the primed inertial coordinate system and keep the well-defined usual (Newtonian) velocity $y^s = dx^s/dt$, s=1,2,3, in the usual inertial coordinate system, we find from Eqs.(2),

$$Y^2 = \delta_{rs} y'^r y'^s = [\frac{c}{2y} \ln \frac{c+y}{c-y}]^2 \delta_{rs} y^r y^s, \; r,s=1,2,3, \tag{6}$$

$$y'^s = [\frac{c}{2y} \ln \frac{c+y}{c-y}] y^s, \; s=1,2,3, \tag{7}$$

and

$$Y = y' = \frac{c}{2} \ln \frac{c+y}{c-y}, \tag{8}$$

where $Y = dX/dT$ is the velocity-length, $y' = (y'^s y'^s)^{1/2}$, $y = (y^s y^s)^{1/2}$, s=1,2,3.

The velocity-space embodied in Eq.(6) is nothing but the so-called Fock velocity-space.

Fock was the first to introduce a velocity-space [5], in order to interpret the Einstein addition law of usual velocities, through defining it in the usual velocity-coordinates $\{y^r\}$, r=1,2,3, with

$$dY^2 = H_{rs}(y) dy^r dy^s, \; r,s=1,2,3, \tag{9a}$$
$$H_{rs}(y) = c^2 \delta^{rs}/(c^2-y^2) + c^2 y^r y^s/(c^2-y^2)^2. \tag{9b}$$

He showed that the geodesic line equation between any two points $y_1^r$ and $y_2^r$ in this velocity-space is linear,

$$y^r = y_1^r + \alpha(y_2^r - y_1^r), \; r=1,2,3, \tag{10}$$
$$0 \leq \alpha \leq 1.$$

Actually, with the standard calculation techniques in Riemann geometry [10], we calculate the Christoffel symbols and find

$$\Gamma_{jk}^i = \begin{cases} 2y^j/(c^2-y^2), & \text{if } i=j=k; \\ y^k/(c^2-y^2), & \text{if } i=j \neq k; \\ y^j/(c^2-y^2), & \text{if } i=k \neq j; \\ 0, & \text{otherwise,} \end{cases} \tag{11}$$

where

$$\Gamma_{ij}^k = H^{km}(y) \Gamma_{ij,m},$$
$$\Gamma_{ij,m} = \frac{1}{2}[\partial H_{im}(y)/\partial y^j + \partial H_{jm}(y)/\partial y^i - \partial H_{ij}(y)/\partial y^m], \; i,j,k,m=1,2,3,$$
$$H^{rs}(y) = (c^2-y^2)\delta^{rs}/c^2 - (c^2-y^2)y^r y^s/c^4.$$

The geodesic line equation is therefore

$$\ddot{y}^r + [2/(c^2-y^2)] \dot{y}^r (y^s \dot{y}^s) = 0, \; r,s=1,2,3, \tag{12}$$

where dot refers to the derivative with respect to velocity-length. Calling new variables

$$w^r = \dot{y}^r/(c^2-y^2), \; r=1,2,3, \tag{13}$$

we are able to rewrite Eq.(12) as

$$\dot{w}^r = 0, \; r=1,2,3. \tag{14}$$

Eqs.(14) have solution of

$$w^r = \text{constant}, \; r=1,2,3. \tag{15}$$

Moreover, due to Eqs.(13) and (14), we have

$$w^s y^r - w^r y^s = \text{constant}, \; r,s=1,2,3. \tag{16}$$

Eqs.(15) and (16) imply linear relations between any two of $y^r$, r=1,2,3. These linear relations can be represented in terms of Eqs.(10).

We evaluate the velocity-length between two points, $y_1^r$ and $y_2^r$, using the integral

$$Y(y_1{}^r, y_2{}^r) = \int_1^2 dY$$

along the geodesic line Eqs.(10), where the lower and upper limits 1 and 2 refer to $y_1{}^r$ and $y_2{}^r$ respectively. At some length, we finally obtain

$$Y(y_1{}^r, y_2{}^r) = \frac{c}{2} \ln \frac{b+a}{b-a}, \tag{17a}$$

$$b = c^2 - y_1{}^r y_2{}^r, \quad r=1,2,3, \tag{17b}$$

$$a = \{(c^2 - y_1{}^i y_1{}^i)(y_2{}^j - y_1{}^j)(y_2{}^j - y_1{}^j) + [y_1{}^k(y_2{}^k - y_1{}^k)]^2\}^{1/2}, \quad i,j,k=1,2,3. \tag{17c}$$

For the special case of $y_1{}^r = 0$ and $y_2{}^r = y^r$, the velocity-length is

$$Y(y^r) = \frac{c}{2} \ln \frac{c+y}{c-y}, \tag{18}$$

where $y = (y^r y^r)^{1/2}$, $r=1,2,3$.

We can re-define the Fock velocity-space in the primed velocity-coordinates $\{y'^r\}$ [11], $r=1,2,3$,

$$dY^2 = \delta_{rs} dy'^r dy'^s, \tag{19}$$

with

$$dy'^r = A^r{}_s(y) dy^s, \tag{20a}$$

$$A^r{}_s(y) = \gamma \delta^{rs} + \gamma(\gamma-1) y^r y^s / y^2, \quad r,s=1,2,3, \tag{20b}$$

where

$$\gamma = 1/(1 - y^2/c^2)^{1/2}. \tag{21}$$

This is simply because

$$\delta_{rs} A^r{}_p(y) A^s{}_q(y) = H_{pq}(y). \tag{22}$$

Certainly, the velocity-length between two points, $y_1'^r$ and $y_2'^r$, $r=1,2,3$, in the Fock velocity-space is

$$Y(y_1'^r, y_2'^r) = [(y_2'^r - y_1'^r)(y_2'^r - y_1'^r)]^{1/2}, \quad r=1,2,3, \tag{23}$$

or

$$Y(y'^r) = (y'^r y'^r)^{1/2}, \quad r=1,2,3, \tag{24}$$

if $y_1'^r = 0$ and $y_2'^r = y'^r$, $r=1,2,3$. Eqs.(18) and (24) together yield Eqs.(6), (7) and (8) provided $y'^r$ and $y^r$, $r=1,2,3$, represent the same point in the Fock velocity-space that judges the statement about the Fock velocity-space embodied in Eq.(6).

The Galilean addition law among primed velocities is linked up with the Einstein addition law among usual velocities [11]. To prove this, we let IFR1 and IFR2 be two inertial frames of reference, and IFR2 move relative to IFR1 with usual velocity $u^r$, $r=1,2,3$, as measured by IFR1. We further let an object move relative to IFR1 and IFR2 with usual velocities $y_1{}^r$ and $y_2{}^r$, $r=1,2,3$, respectively, as measured by IFR1 and IFR2. Three primed velocities are $y_1'^r$, $y_2'^r$, $u'^r$, $r=1,2,3$, respectively corresponding to $y_1{}^r$, $y_2{}^r$, $u^r$.

The Galilean addition law among the three primed velocities reads

$$y_2'^r = y_1'^r - u'^r, \quad r=1,2,3. \tag{25}$$

Using Eqs.(17) and (23), we have for IFR1

$$[(y_1'^r - u'^r)(y_1'^r - u'^r)]^{1/2} = \frac{c}{2} \ln \frac{b+a}{b-a} \equiv \rho_1$$

with

$$b = c^2 - y_1{}^i u^i,$$

$$a = \{(c^2 - u^i u^i)(y_1{}^j - u^j)(y_1{}^j - u^j) + [u^k(y_1{}^k - u^k)]^2\}^{1/2},$$

or equivalently,

$$c^2 \tanh^2(\rho_1/c) = c^2\{(c^2 - u^i u^i)(y_1{}^j - u^j)(y_1{}^j - u^j) + [u^k(y_1{}^k - u^k)]^2\}/(c^2 - u^i u^i)^2. \tag{26}$$

Using Eqs.(18) and (24), we have for IFR2

$$(y_2'^r y_2'^r)^{1/2} = \frac{c}{2} \ln[(c+y_2)/(c-y_2)] \equiv \rho_2,$$

or equivalently,

$$y_2{}^2 = c^2 \tanh^2(\rho_2/c). \tag{27}$$

Eq.(25) or $\rho_1=\rho_2$ combines with Eqs.(26) and (27) to give rise to the Einstein addition law among $y_1^r$, $y_2^r$ and $u^r$,

$$y_2^r = \sqrt{1-\frac{u^2}{c^2}}\ \{(y_1^r-u^r)+(\frac{1}{\sqrt{1-\frac{u^2}{c^2}}}-1)u^r\frac{u^s(y_1^s-u^s)}{u^2}\}/[1-\frac{y_1^k u^k}{c^2}],\ r,s,k=1,2,3. \qquad (28)$$

\* \* \*

Eqs.(6) and (19) convince us of validity of the classical Maxwell velocity and velocity rate distributions for free particles in the primed velocity-coordinates $\{y'^r\}$, r=1,2,3. They are

$$P(y'^1,y'^2,y'^3)dy'^1dy'^2dy'^3 = N(\frac{m}{2\pi K_B T})^{3/2}\exp[-\frac{m}{2K_B T}(y')^2]dy'^1dy'^2dy'^3 \qquad (29)$$

and

$$P(y')dy' = 4\pi N(\frac{m}{2\pi K_B T})^{3/2}(y')^2\exp[-\frac{m}{2K_B T}(y')^2]dy', \qquad (30)$$

where $y'=(y'^r y'^r)^{1/2}$, r=1,2,3, Using Eqs.(7), (8) and (20) to represent these distributions in the usual velocity-coordinates $\{y^r\}$, r=1,2,3, we find

$$P(y^1,y^2,y^3)dy^1 dy^2 dy^3 = N\frac{(m/2\pi K_B T)^{3/2}}{(1-y^2/c^2)^2}\exp[-\frac{mc^2}{8K_B T}(\ln\frac{c+y}{c-y})^2]dy^1 dy^2 dy^3 \qquad (31)$$

and

$$P(y)dy = \pi c^2 N\frac{(m/2\pi K_B T)^{3/2}}{(1-y^2/c^2)}(\ln\frac{c+y}{c-y})^2\exp[-\frac{mc^2}{8K_B T}(\ln\frac{c+y}{c-y})^2]dy \qquad (32)$$

Both distribution functions $P(y^1,y^2,y^3)$ in Eq.(31) and $P(y)$ in Eq.(32) reduce to their previous non-relativistic formulas [12],

$$N(\frac{m}{2\pi K_B T})^{3/2}\exp[-\frac{m}{2K_B T}(y)^2] \qquad (33)$$

and

$$4\pi N(\frac{m}{2\pi K_B T})^{3/2}(y)^2\exp[-\frac{m}{2K_B T}(y)^2], \qquad (34)$$

respectively when y is small enough.

It is the characteristic of distribution functions $P(y^1,y^2,y^3)$ and $P(y)$ that they go to zero as y approaches speed c. This is consistent with the spirit of the modified special relativity theory. The deviations of $P(y^1,y^2,y^3)$ from Eq.(33) and of $P(y)$ from Eq.(34) might be testable in the molecular or atom or electron beam experiments and provide us with an evidence for the local structures Eqs.(2) of gravity-free space and time in the usual inertial coordinate system.

ACKNOWLEDGMENT
The author greatly appreciates the teachings of Prof. Wo-Te Shen. The author thanks Prof. Mark Y. Mott and Dr. Allen E. Baumann for helpful suggestions.